\documentclass{svproc}

\usepackage{booktabs}
\usepackage{footmisc}
\usepackage{colortbl}
\usepackage{pifont}
\usepackage{graphicx}
\usepackage{mathrsfs}
\usepackage{amsmath}
\usepackage{mathtools}
\usepackage{amsfonts}
\usepackage[square,numbers]{natbib}
\usepackage{vmargin}
\usepackage{float}
\usepackage{tikz}
\usepackage{lscape}
\usepackage{subfig}
\usepackage{multicol}

\newcommand{\rmd}{ {\rm d}}

\newcommand\blfootnote[1]{%
  \begingroup
  \renewcommand\thefootnote{}\footnote{#1}%
  \addtocounter{footnote}{-1}%
  \endgroup
}

\begin{document}
\mainmatter

\title{Consistent Approximation of Epidemic Dynamics on Degree-heterogeneous Clustered Networks
}
\titlerunning{Approximate epidemic dynamics on heterogeneous clustered networks}

\author{A.~Bishop\inst{1} \and I.~Z.~Kiss\inst{2} \and T.~House\inst{3}}

\authorrunning{A.~Bishop et al.} 

\tocauthor{Alex Bishop, Istvan Z. Kiss, and Thomas House}

\institute{Centre for Complexity Science, University of Warwick, Coventry, CV4 7AL, UK\\
\email{A.Bishop@warwick.ac.uk}\\
\and
Department of Mathematics, School of Mathematical and Physical Sciences, University of Sussex, Brighton, BN1 9QH\\
\email{I.Z.Kiss@sussex.ac.uk}
\and
School of Mathematics, University of Manchester, Manchester, M13 9PL, UK.\\
\email{thomas.house@manchester.ac.uk}}

\maketitle

\begin{abstract}
\noindent{}Realistic human contact networks capable of spreading infectious
disease, for example studied in social contact surveys, exhibit both significant
degree heterogeneity and clustering, both of which greatly affect epidemic
dynamics. To understand the joint effects of these two network properties on epidemic
dynamics, the effective degree model of Lindquist et al.~\cite{Lindquist} is reformulated with
a new moment closure to apply to highly clustered networks. A simulation study
comparing alternative ODE models and stochastic simulations is performed for
SIR (Susceptible--Infected--Removed) epidemic dynamics, including a test for
the conjectured error behaviour in \cite{Pellis:2015}, providing evidence that
this novel model can be a more accurate approximation to epidemic dynamics
on complex networks than existing approaches.
\end{abstract}

\section{Introduction\label{sec:intro}}

\blfootnote{Work supported by the Engineering and Physical Sciences Research Council Grant numbers EP/I01358X/1 and EP/N033701/1.}


Networks offer an unprecedented opportunity to represent and model contacts
between interacting units at all scales ranging from proteins and individuals
to countries via air transportation. This additional degree of freedom has lead
to extensive modelling and analysis in mathematical epidemiology and has
allowed the development of a number of network-based models, which are either
based or parametrised by real network data, if available, or by using synthetic
or theoretical networks which reflect and reproduce some observed local or
global properties of real-world networks
\citep{keeling2005networks,danon2011networks,pastor2014epidemic,kiss2017mathematics}.
Idealised networks often offer a greater degree of analytical tractability,
which in turn offers clearer insight into the impact of network properties on
epidemic outbreak threshold, final epidemic size or prevalence and on
effectiveness or choice of control measures.

The many degrees of freedom offered by networks come at the cost of
computational and mathematical complexity.  In order to handle this, and for a
systematic investigation and understanding of network processes, it is
desirable not to rely solely on stochastic simulations. As a result a number of
differential equation-based models have been developed including pairwise
\citep{rand2009,Keeling1999,house2011,taylor2012markovian}, PGF or edge-based
compartmental models \citep{miller2011edge,miller:ebcm_overview} and effective
degree models \citep{Lindquist,gleeson2013binary}, to name just a few. The goal
of all these models is the same: to derive a set of low-dimensional system of
ordinary differential equations (ODEs) where variables correspond to some
average quantity from the stochastic process---e.g.\ expected prevalence---
over time. All such mean-field models require choice of a `state space'. For
example, pairwise models initially concentrate on the expected number of nodes
in different states, while effective degree models concentrate on the expected
number of star like structures, i.e.\ a central node with all its neighbours,
and the  possible state that these can be in. Once chosen, evolution equations
for these variables are derived.  This, often heuristic, step still involves a
precise book-keeping which in general yields a dependency on higher order
states or moments, e.g.\ for pairwise models the expected number of infected
nodes depends on the expected number of edges where one node is infected and
the other is susceptible. These newly introduced higher-order structures or
moments require further equations and and hence to curtail the dependency on
higher-order moments and the fast growth in the number of equations `closures'
are needed. This amounts to approximating higher-order moments in  terms of
lower order ones. The performance of mean-field models is then often tested by
direct comparison to results based on explicit stochastic network simulations.
If the process is successful and the mean-field model works well, the analysis
of the stochastic epidemic on networks is mapped into analysing a system of
ODEs.  This can be done using dynamical systems tools and such analysis often
leads to analytical results which explicitly reveal the interaction between
network and disease characteristics. More importantly, it shows how the
fundamental properties of the network impact on growth rate, epidemic
threshold, final epidemic size and so on.

Degree or contact heterogeneity and degree-based mixing is well accounted for
in existing differential equation models, but epidemics on networks which are
clustered (i.e.\ where two nodes with a common neighbour are highly likely to
be neigbours of each other \cite{watts1998collective}) pose more of a
challenge.  A major factor of this difficulty is the non-unique way in which
global or network-level clustering can be achieved, and the same level of
clustering can be achieved while keeping the degree distribution the same but
using different distributions or different sets of motifs
\citep{newman2003properties,kiss2008comment,green2010large,ritchie2014higher}. 

In general, clustered bond percolation-type 
\citep{gleeson2009bond,newman2009random, miller2009percolation,karrerclust2010}
or PGF-based models \citep{vmclust,ritchie2014beyond, house2011} consider specific forms 
of clustering (e.g.\ non-overlapping triangles) while pairwise models 
\citep{Keeling1999,house2009motif,wrap5586,Keeling:2016, house2011} usually work best in
the case where clustering is `randomly' distributed, as is the case when using 
rewiring algorithms such as the big-V 
\cite{house2011insights,bansal2009exploring,green2010large}, which precludes 
certain kinds of interaction between clustering and degree heterogeneity. The 
question of how to approximate epidemic dynamics on a more general complex 
network remains open, and is the topic of this paper.

In general, for clustered networks it is difficult to derive accurate
differential equation models for epidemics. In this paper, we present an
approach based on generalisation of the effective degree model to clustered
networks, and we show that this newly-derived model outperforms the
state-of-the-art models and display excellent agreement with results based on
stochastic simulations for a range of degree distributions and clustering
values. 

The paper is structured as follows: In Sect. \ref{sec:networks} an overview
of relevant complex network concepts and terminology are presented; Sect.
\ref{sec:net_gen} introduces methods of generating clustered networks with a
specific degree distribution; in Sect. \ref{sec:tree_dyn} existing ODE models
are introduced and discussed; Sect. \ref{sec:clus_dyn} presents the extension
of ODE models to dynamics on clustered networks including our novel ODE model;
Sect. \ref{sec:sim_stud} details the simulation study performed to compare
between models and test the closure performance against conjectured errors.

\section{Model Definitions\label{sec:networks}}

\subsection{Networks}

A \textit{network} (or \textit{graph}) is a pair $\mathcal{G} = (\mathcal{N},
\mathcal{L})$ where $\mathcal{N}$ is a size-$N$ set of \textit{nodes} and
$\mathcal{L}\subseteq \mathcal{N} \times \mathcal{N}$ is a set of
\textit{links}. The information about a network can be usefully encoded in
terms of an \textit{adjacency matrix}, which has elements $G_{ab}=1$ if
$(a,b)\in \mathcal{L}$, and zero otherwise.

\noindent We will consider simple undirected networks without self edges meaning that
$G_{aa} = 0$ and $G_{ab} = G_{ba}$, $\forall a, b \in \mathcal{N}$.  The
\textit{degree} of node $a$ is the number of links that it participates in,
i.e.
\begin{equation}
	k_a := \sum_{b\in\mathcal{N}} G_{ab} \text{ .}	
\end{equation}
We assume that all degrees are integers between $1$ and the \textit{maximum degree} $M$.
We will use angled brackets to refer to mean values of functions of degree
across the network, i.e.\ for an arbitrary function $f$,
\begin{equation}
	\langle f(k) \rangle := \frac{1}{N} \sum_{a\in\mathcal{N}} f(k_a) \text{ .}	
\end{equation}
One particularly important such expectation is the \textit{probability generating
function} (PGF) which is $\psi(x) := \langle x^k \rangle $.

\noindent A final network property of interest in this work is the
\textit{clustering coefficient},
\begin{equation}
	\phi := \frac{\sum_{a,b,c\in\mathcal{N}} G_{ab}G_{bc}G_{ca}}%
	{\sum_{a,b,c\neq a\in\mathcal{N}} G_{ab}G_{bc}} \in [0,1] \text{ .}	
\end{equation}
Our interest is in networks with large size ($N\gg 1$), degree distributions
with a non-infinite variance ($\langle k \rangle^2 \leq \langle k^2\rangle <
\infty$), and clustering coefficients that can be non-zero but are not
particularly large ($\phi \in [0,0.3]$).

\subsection{Epidemic Dynamics}

We will consider SIR (Susceptible--Infective--Removed) dynamics. At the
individual level, an individual $a$ has a random state $X_a \in\{S,I,R\}$.
Individuals in state $I$ move to state $R$ at rate $\gamma$, and individuals in
state $S$ move to state $I$ at a rate $\tau$ multiplied by the number of $I$
individuals they are linked to on the network.

At the population level, we will consider expected total node, pair and triple
counts in given disease state configurations,
\begin{align}
  \label{eqn:singles}
	[A] =&\; \mathbb{E} \sum_{a\in\mathcal{N}} \mathbf{1}_{\{X_a = A\}} \text{ ,}\\ 
  \label{eqn:doubles}
	[AB] =&\; \mathbb{E} \sum_{a,b\in\mathcal{N}} \mathbf{1}_{\{X_a = A \& X_b = B\}}
	G_{ab} \text{ ,}\\ 
  \label{eqn:triples}
	[ABC] =&\; \mathbb{E} \sum_{a,b,c\neq a\in\mathcal{N}}
	\mathbf{1}_{\{X_a = A \& X_b = B \& X_c = C\}} G_{ab} G_{bc} \text{ ,}
\end{align} 
where $A,B,C \in \{S,I,R\}$ and $\mathbf{1}$ is the indicator function. We will
distinguish notationally between closed and open triples,
\begin{align}
	[ABC]_{\triangle} =&\; \mathbb{E} \sum_{a,b,c\in\mathcal{N}}
	\mathbf{1}_{\{X_a = A \& X_b = B \& X_c = C\}} G_{ab} G_{bc} G_{ca} \text{ ,}\\ 
	[ABC]_{\wedge} =&\; [ABC] - [ABC]_{\triangle} \text{ ,}
\end{align} 
which will be particularly important later. We will also consider more detailed
states---in particular $[A_{s,i}]$ represents the expected number of nodes in
state $A$ with $s$ susceptible and $i$ infective contacts.  Another important
definition is the \textit{correlation} between states 
\begin{equation}
	\mathcal{C}_{AB} = \frac{N}{\langle k \rangle} \frac{[AB]}{[A][B]} \text{ ,}
\end{equation}
which expresses how much more likely an $[AB]$ edge is over the null model. The understanding of this is key to the role of networks in shaping epidemic
dynamics \citep{keeling1999pair}.

Our aim in this paper is to find methods for approximation of the expected
population-level behaviour of the epidemic dynamics that are, as much as
possible, logically consistent, well motivated, and accurate.

\section{Models of Network Generation\label{sec:net_gen}}

Typically, a full epidemic network is not directly available from data and so a
standard method is to work with probabilistic models for the network that
respect certain observable summary statistics---in our case, the degree
distribution and clustering coefficient. We will now present several such
algorithms in outline form---more detail on these can be found in relevant
papers and textbooks, e.g.\ Newman \cite{newman2009networks}.

\begin{figure}[H]
\begin{center}
\includegraphics[width=0.75\textwidth]{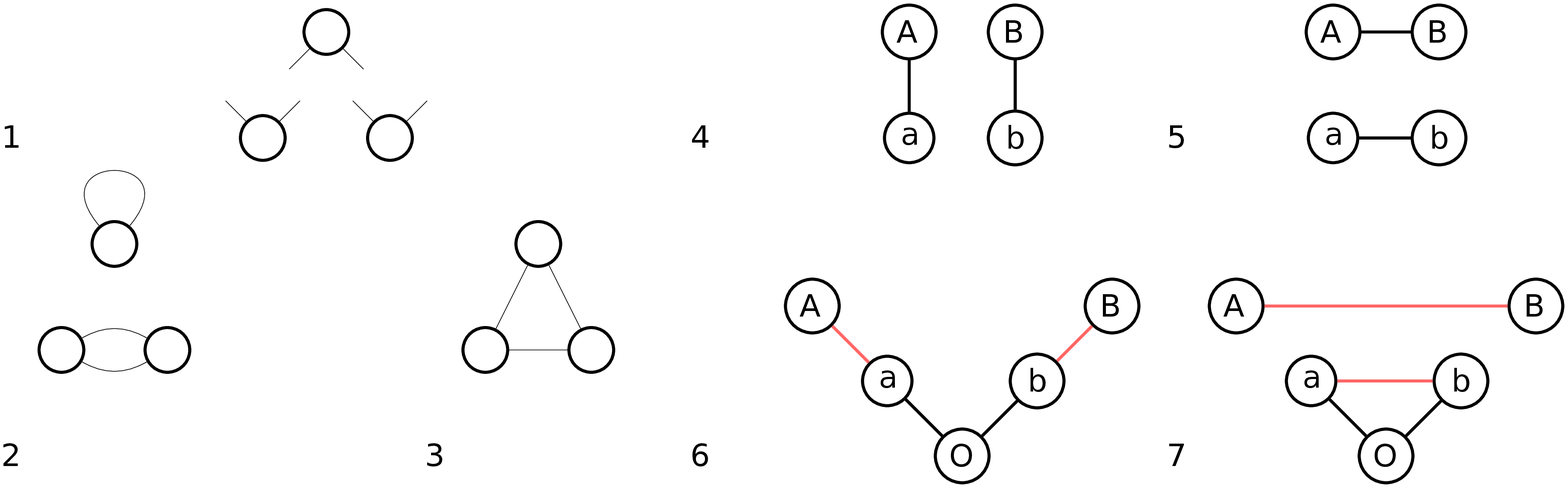}
\caption{\label{fig:bigV} The configuration model starts with a set of nodes
with half-edges (\ding{172}) and constructs a network by pairing these up to
form a full-edge at random, resulting in configurations like \ding{173} \&
\ding{174}.\\ \ding{173} shows an example where the configuration model can
lead to self and duplicate edges, where as \ding{174} gives a valid
configuration. \\ \ding{175} $\to$ \ding{176} will rewire a network such that
local structure is lost, for example performing many of these edge swaps yields
a network with both neglibile clustering and degree assortativity. In addition
this rewiring introduces the small world property to a lattice.\\ The
clustering of a network may be increased by looking for V-shaped configurations
(\ding{177}) and performing the rewiring \ding{177} $\to$ \ding{178} if it
increases the overall clustering coefficient of the network.\\
}
\end{center}
\end{figure}

In $n$-regular networks, all nodes have the same degree i.e.\ $k_a = n, \forall
a\in\mathcal{N}$. A typical such network can be generated by starting with a
network that is atypical but assuredly $n$-regular, for example a
one-dimensional $(n/2)$-nearest neighbour. The network is then rewired by
removing two edges $(a,b)$ and $(c,d)$ sampled (without replacement) uniformly
at random from $\mathcal{L}$, and then adding new edges $(a,c)$ and $(b,d)$
(Fig. \ref{fig:bigV} \ding{174} $\rightarrow$ \ding{175}).  Performing a
large number of such `edge swaps' will result in a $n$-regular network that is
representative of this class of graphs.

Erd\H{o}s-R\'{e}nyi networks \cite{gilbert1959}, often referred to simply as
`random graphs' due to their importance, are defined such that each possible
link is present with independent probability $q$. This leads to a binomial
degree distribution although practically this will always be very well
approximated by a Poisson distribution with mean $\langle k \rangle = (N-1) q$.

In the configuration model (CM), nodes are given a number of `stubs', which are
then paired uniformly at random \cite{molloy1995critical}. This construction is
useful for the development of asymptotic results, but can cause problems for a
given finite value of $N$ due to the presence of `defects'---self-edges,
multiple links between nodes, and stubs that are not eventually paired with
others.

Networks with a given degree distribution can be generated following the
procedure described in \cite{del2010efficient}, which generates a sample of a
given degree distribution that is graphical (meaning that a network with this
degree distribution can be constructed without defects such as self or multiple
edges) and then directly samples a statistically independent network.  This
algorithm is better behaved than the Configuration Model as it always produces
a simple graph without defects, backtracking or rejections.

Algorithms that generate clustered heterogeneous networks using one random
process typically generate special network topologies. We therefore generate
clustered networks via a two-step process in which we first generate a network
of the required degree distribution, and then increase the clustering
coefficient using a rewiring method known as the `big-V'
\citep{house2011insights,bansal2009exploring,green2010large}. This involves finding a `V'
configuration in the network (Fig. \ref{fig:bigV} \ding{177}) and
proposing a rewiring which produces a triangle and a separate edge (Fig.
\ref{fig:bigV} \ding{178}).

This rewiring is then performed provided the clustering coefficient is
increased. This method preserves the degree distribution whilst increasing the
clustering coefficient (empirically) up to between $\phi=0.3$ and $\phi=0.4$
before the acceptance ratio of proposed moves critically slows down.  The
origin node, $c$, is selected with a weight proportional to $k_c(k_c-1)$ so
that the expected proportion of possible triangles present around each node
does not depend on its degree \cite{Cbias}.

\section{Epidemic Dynamics on Locally Tree-like Networks\label{sec:tree_dyn}}

In the limit as $\phi \rightarrow 0$ while $N \rightarrow \infty$, it is
possible to obtain results for the expected population-level epidemic dynamics
that are known to be asymptotically exact.

\subsection{Simple SIR Model\label{ssec:simpleSIR}}

It is possible to show \citep{lumpingpaper} that the following equations hold
for an arbitrary network:
\begin{equation}
  \label{eqn:pairS}
	\frac{\rmd }{\rmd t}[S] =  - \tau [SI] \text{ ,} \qquad
	\frac{\rmd }{\rmd t}[I] =  \tau [SI] - \gamma [I] \text{ .} 
\end{equation}
There are two limits in which we can accurately approximate the pair variable
$[SI]$ in terms of the node variables: (i) for an $n$-regular graph, as
$n\rightarrow N-1$; and (ii) for an ER graph with mean degree $\langle k
\rangle$. In each case we take $[SI] \approx \langle k \rangle [S][I]/N$
with $\beta := \tau \langle k \rangle$ to end up with a special case of the
classic model in mathematical epidemiology introduced almost a century ago by
Kermack and McKendrick \cite{kermack1927and}, often called the simple SIR
model, 
\begin{equation}
  \label{eqn:SIR}
	\frac{\rmd }{\rmd t}[S] =  - \frac{\beta}{N} [S][I] \text{ ,} \qquad
	\frac{\rmd }{\rmd t}[I] =  \frac{\beta}{N} [S][I] - \gamma [I] \text{ .} 
\end{equation}
Homogeneous mixing in a population is a poor assumption and more structural
information is required to focus on the underlying network of contacts.

\subsection{Pairwise Model\label{ssec:PW}}

The pairwise model was one of the first steps toward this more realistic
contact structure, and primarily concerns $n$-regular graphs. In this approach,
we continue to write down equations in the form~\eqref{eqn:pairS},
\begin{align}
  \label{eqn:pairSS}
  \frac{\rmd }{\rmd t}[SS] =&\;  - 2\tau [ISS], \\
  \label{eqn:pairSI}
  \frac{\rmd }{\rmd t}[SI] =&\;  \tau \Big([ISS] - [ISI] - [IS]\Big) - \gamma [IS],\\
  \label{eqn:pairII}
  \frac{\rmd }{\rmd t}[II] =&\;  2\tau \Big([ISI] + [IS]\Big) - 2\gamma [II] \;.
\end{align}
The closed pairwise model \cite{keeling1999pair,rand1999pair} is gained by
taking  the unclosed ODEs (\ref{eqn:pairS} and
\ref{eqn:pairSS}--\ref{eqn:pairII}) and approximating the number of triples,
$[ABC]$, in the system using a moment closure originally attributed to
Kirkwood \cite{kirkwood1942},
\begin{align}
  \label{eqn:kirkclus}
  [ABC] \approx 
	\frac{n-1}{n} \frac{[AB][BC]}{[B]} \;.
\end{align}

\subsection{Effective Degree (ED) Model\label{ssec:ED}}

Ball and Neal \cite{ballneal2008} introduced the notion of an effective degree. In the Ball
and Neal construction, individuals (nodes) begin with a number of unpaired
half-links/stubs---an \textit{effective degree}---and a contact network is
constructed along with SIR epidemic dynamics. 

The dynamics occur as follows: stubs of infected nodes randomly connect with
stubs of susceptible nodes at rate $\tau$, reducing the effective degree of
both nodes by one and infecting the susceptible node; infected nodes become
recovered at rate $\gamma$, at which point they connect their remaining stubs
uniformly at random to any remaining unpaired stubs in the network thus
reducing their effective degree to zero.  This process results in a
configuration model network \citep{molloy1995critical}.

Following Ball and Neal \cite{ballneal2008}, Lindquist et
al.\cite{lindquist2011ED} developed an effective degree model categorising each
node by its own disease state (S, I or R) along with the number of neighbours
in each disease state resulting in the network being separated into classes
representing the state of a node and its neighbours. For example, $S_{s,i}$
denotes the star motif corresponding to a susceptible node where $s$ and $i$
are the number of susceptible and infected neighbours respectively.

Analogously to the Pairwise model (see Sect. \ref{ssec:PW}), ODEs describing the
time evolution of network motifs may be written down; however for the Effective
Degree model there are $M(M+3)$ equations representing the time evolution of
each of the possible star motifs (where $M$ is the maximum degree) compared to
the equations of the pairwise model (\ref{eqn:pairSS}-\ref{eqn:pairII}) which
represent the time evolution of all node and all edge combinations, of which
there are $5$ ($[S], [I], [SS], [SI], [II]$).

The susceptible node at the centre of a $S_{s,i}$ star experiences a force of
infection, $\tau$, from each of its $i$ infected neighbours with infection
resulting in the transition from an $S_{s,i}$ star to an $I_{s,i}$ star.  Each
of the $i$ infectious neighbours recover at rate $\gamma$ and thus transition
from the $S_{s,i}$ class to the $S_{s,i-1}$ class at rate $\gamma i [S_{s,i}$].
By the same reasoning the rate at which transitions from $S_{s,i+1}$ to
$S_{s,i}$ occur is $\gamma (i+1) [S_{s,i+1}]$ .  Infection of one the $s$
susceptible neighbours results in a transition from $S_{s,i}$ to $S_{s-1,i+1}$
occurring at a rate of
\begin{equation}
  \dfrac{\sum_{j,l} \tau jl [S_{j,l}]}{\sum_{j,l}j[S_{j,l}]} s [S_{s,i}]\text{ ,}
  \label{eqn:ISSrate}
\end{equation}
where above and henceforth notation of the form ${\sum_{j,l} =
\sum^M_{k=1}~\sum_{j+l=k}}$ is adopted to aid brevity.  The rate derives from
the fact that new infections are generated at rate $ \sum_{j,l} \tau l [S_{j,l}]$
which in turn causes the effective degree of the susceptible neighbours of the
now infected $[S_{j,l}]$ to change at rate $ \sum_{j,l} \tau j l [S_{j,l}]$. Put in
the notation of Sect. \ref{ssec:PW} this can be expressed as the rate at
which transitions $[ISS] \rightarrow [IIS]$ occur which is $\tau [ISS]$ where,
\begin{equation}
  [ISS] = \sum_{j,l} [IS_{j,l}S] = \sum_{j,l}jl[S_{j,l}]\text{ .}
  \label{}
\end{equation}
As we wish to express the rate at which a susceptible neighbour of a
$S_{s,i}$ star becomes infected, i.e.\ the rate of $[ISS_{s,i}] \rightarrow
[IIS_{s-1,i+1}]$,
we must account for the probability that the $S$--$S$ link in the $[I S
S]$ triple will connect to a $S_{s,i}$ star which is given by the following
identity,

\begin{equation}
  [ISA_{s,i}] = [ISA] \dfrac{s[A_{s,i}]}{\sum_{j,l}jA_{j,l}}, \; A \in \{S,I\}. 
  \label{eqn:end_ident}
\end{equation}

Putting this all together then yields the rate given in (\ref{eqn:ISSrate}).
{Mutatis mutandis} the rate of transition from $S_{s+1,i-1}$ to
$S_{s,i}$ is obtained.

\begin{figure}
\begin{center}
\includegraphics[width=0.5\textwidth]{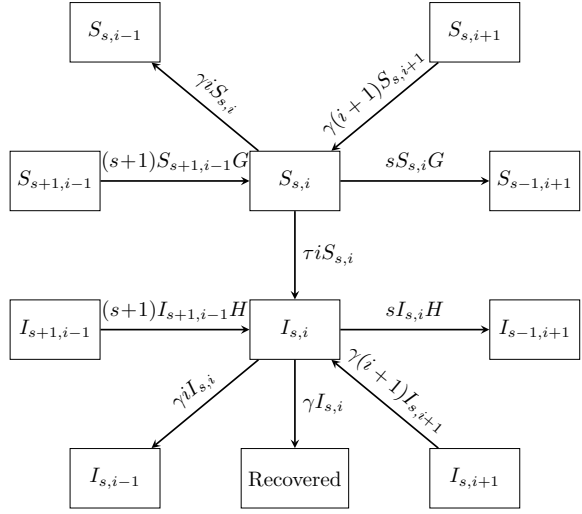}
\caption{\label{fig:EDflow} Flow chart for the effective degree model
illustrating the flow rates in to and out of motifs $S_{s,i}$ and $I_{s,i}$,
where $G= \frac{\sum_{j,l} \tau jl [S_{j,l}]}{\sum_{j,l}j [S_{j,l}]}$, and $H=
\frac{\sum_{j,l} \tau l^2 [S_{j,l}]}{\sum_{j,l}j [I_{j,l}]}$. Reproduced from
\citep{lindquist2011ED}.}
\end{center}
\end{figure}

Using the above rates the set of ODEs describing the time evolution of star
motifs with a central susceptible are obtained  as \eqref{eqn:EDs} below.
Following similar reasoning allows \eqref{eqn:EDi} below to be derived
with the extra term $\gamma [I_{s,i}]$ corresponding to the recovery of the
infectious individual at the centre of the star:
\begin{align}
  \label{eqn:EDs}
  \frac{\rmd [{S}_{s,i}]}{\rmd t} =&  \gamma \Big( (i+1) [S_{s,i+1}] - i [S_{s,i}] \Big) + \frac{ \sum_{j,l} \tau j l [S_{j,l}] }{ \sum_{j,l} j [S_{j,l}]} \Big( (s+1) [ S_{s+1,i-1} ] - s [ S_{s,i} ] \Big) - \tau i [ S_{s,i} ] \\
  \label{eqn:EDi}
  \frac{\rmd [{I}_{s,i}]}{\rmd t} =& \gamma \Big( (i+1) [ I_{s,i+1} ] - i [ I_{s,i} ] \Big) + \frac{ \sum_{j,l} \tau l^2 [S_{j,l}] }{ \sum_{j,l} j [I_{j,l}]} \Big( (s+1) [ I_{s+1,i-1} ] - s [ I_{s,i} ] \Big) - \gamma [ I_{s,i} ] + \tau i [ S_{s,i} ] 
\end{align}
The constraints upon $s$ and $i$ are given by $\left\{(s,i) : s \geq 0 ,\, i
\geq 0 ,\, s+i \leq M \right\} $.  
Figure \ref{fig:EDflow} summarises these two sets of equations governing the system in graphical form.

It is noteworthy to remark that the closure in this model is more implicit than
that of the pairwise model, i.e.\ rather than approximating higher order states
with lower order states one makes the assumption that the infectious pressure
on the susceptible neighbours of the central node is equal to the population
average.

\subsection{Probability Generating Function (PGF) Methods}

PGF models~\citep{Volz:2008,Miller:2012} provide low-dimensional
representations of epidemic dynamics on configuration models and have been
proved to be asymptotically exact
\citep{Decreusefond:2012,Bohman:2012,Barbour:2013,Janson:2014}.

In the simplest form of PGF model, given in \cite{Miller:2011}, epidemic
dynamics can be captured by one differential equation,
\begin{equation}
	\frac{\rmd \theta}{\rmd t} = - \tau \theta + \gamma (1-\theta)
	+ \tau \frac{\psi'(\theta)}{\psi'(1)} \text{ .}
\end{equation}
The Volz variable $\theta$ represents the probability that a `test node' with
one link remains susceptible, and its use is responsible for the massive
simplicity of this model.

\section{Dynamics on Clustered Networks\label{sec:clus_dyn}}

\subsection{Pairwise Model\label{ssec:clustPW}}

The Kirkwood closure of Sect. \ref{ssec:PW} can be extended to clustered networks \cite{keeling1999pair} by introducing a correlation term, {\allowbreak $\mathcal{C}_{CA} = \tfrac{N}{\langle k \rangle} \tfrac{[CA]}{[A][C]}$}, that accounts for the number of transitive links between A and C by measuring the observed $[CA]$ compared to the number of A--C pairs one would expect given independence between C and A, $ \tfrac{\langle k \rangle [A][C]}{N}$. If $\mathcal{C}_{CA}=1$ then nodes of type C and nodes of type A are connected at random. Given a clustering coefficient, $\phi$, the clustered Kirkwood closure is,
\begin{align}
  \label{eqn:kirkclus}
  [ABC] \approx 
	\frac{n-1}{n} \frac{[AB][BC]}{[B]} \left( (1-\phi)  + \phi\, \mathcal{C}_{AC} \right) \;.
\end{align}

\subsection{Clustered PGF}

House and Keeling \cite{house2011insights} introduced the clustered PGF model, which consists of
the following differential equations with $[S] = N \psi(\theta)$ and $Y = \sum_k k [I_k]$:
\begin{equation}
\frac{\rmd \theta}{\rmd t}  = - \tau \frac{[SI]}{N \psi'(\theta)}\; ,\quad
\frac{\rmd [I]}{\rmd t}  = \tau [SI] - \gamma [I] \; ,\quad
\frac{\rmd Y}{\rmd t}   = \tau \left( \frac{\theta \psi''(\theta)}{\psi'(\theta)}
 + 1 \right) - \gamma Y \; ,
\end{equation}
together with \eqref{eqn:pairSS}, \eqref{eqn:pairSI} and \eqref{eqn:pairII}, 
and the closures
\begin{align}
	[ISS] & \approx 
  \frac{[SI][SS]\psi''(\theta)}{N\psi'(\theta)^2} \left( (1-\phi)  + \phi \langle k \rangle 
	\frac{[SI]}{N^2 \theta \psi'(\theta)^2} \right) \; .
\\
 [ISI] & \approx \frac{[SI]^2\psi''(\theta)}{N\psi'(\theta)^2} \left( (1-\phi)  
 + \phi \langle k \rangle N \frac{[II]}{{Y}^2} \right) \; .
\end{align}
The clustered PGF is the main model that attempts to deal with epidemics on
clustered heterogeneous networks and as such is the main direct comparator to
our new approach. 

\subsection{A New Model}

The ED degree model (Sect. \ref{ssec:ED}) approximates the epidemic
dynamics on a network of arbitrary degree distribution extremely well; however,
in the large network limit the clustering coefficient of the configuration
model network it explains tends to zero, which does not realistically explain
real world contact networks. A new set of equations extending the effective
degree model to clustered networks is now presented.

\begin{figure}[H]
\begin{center}
\includegraphics[width=0.3\textwidth]{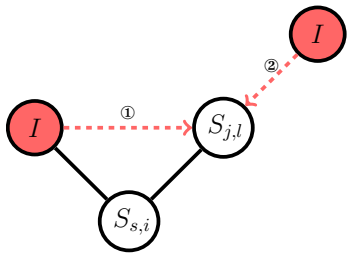}
\end{center}
\caption{\label{fig:inout} Illustration of transmission events occuring within
the neighbourhood of a central node ($S_{s,i}$) and outside of it. The edge
labelled \ding{172} denotes a transmission event within the neighbourhood and
the edge labelled \ding{173} denotes a transmission event outside of the
neighbourhood of the central $S_{s,i}$.} 
\end{figure}

An exact version of the ED model may be written as follows,
\begin{align}
  \label{eqn:exactED_S} \frac{\rmd [{S}_{s,i}]}{\rmd t} =& - \tau i [ S_{s,i} ] + \gamma \Big( (i\!+\!1) [ S_{s,i+1} ] - i [ S_{s,i} ] \Big) -  \tau [{ISS_{s,i}}]_{\bigtriangleup} - \tau [ISS_{s,i}]_{\wedge} + \\& \tau [{ISS_{s+1,i-1}}]_{\triangle} \nonumber + \tau [ISS_{s+1,i-1}]_{\wedge}\; , \\
  \label{eqn:exactED_I} \frac{\rmd [{I}_{s,i}]}{\rmd t} =&  \tau i [ S_{s,i} ]  + \gamma \Big( (i\!+\!1) [ I_{s,i+1} ] - i [ I_{s,i} ] \Big) - \tau [{ISI_{s,i}}]_{\triangle} - \tau [ISI_{s,i}]_{\wedge} + \tau [{ISI_{s+1,i-1}}]_{\triangle} \nonumber \\& + \tau [ISI_{s+1,i-1}]_{\wedge} - \gamma [ I_{s,i} ] + \tau \Big( (s\!+\!1) [ I_{s+1,i-1} ] - s [ I_{s,i} ] \Big)\; ,
\end{align}
where we separate out transmission events happening within and outside of the
neighbourhood of the star: $[ISS_{s,i}]_{\triangle}$ denotes a closed triple
forming a triangle (e.g. edge \ding{172} Fig. \ref{fig:inout}) and
$[ISS_{s,i}]_{\wedge}$ denotes an unclosed triple (e.g. edge \ding{173} Fig.
\ref{fig:inout}). 

\begin{table}[H]
\def\arraystretch{2.0}
\begin{center}
\caption{Counts of possible triangles around a node in state $A_{s,i}$. \label{tab:SItris}}
{\setlength{\tabcolsep}{1em}
\begin{tabular}{| c || c | c | c |}
  \hline
  \textbf{Triangle states} & $[SSA_{s,i}]_{\triangle}$ & $[ISA_{s,i}]_{\triangle}$
 	& $[IIA_{s,i}]_{\triangle}$\\
  \hline
  \textbf{Combinations} & $\tfrac{1}{2}s(s-1)$ & $is$ & $\tfrac{1}{2}i(i-1)$\\
  \hline
\end{tabular}}
\end{center}
\end{table}

In a network with a clustering coefficient of zero then $[ABC]_{\triangle}=0$
and a closure on the $[ABC]_{\wedge}$ terms will yield the ED
model \cite{lindquist2011ED}; however an effective closure for clustered
networks must be made on both $\triangle$ and $\wedge$ terms.

Given a clustering coefficient, $\phi$, there are an expected $\phi k(k-1)/2$
triangles around a given degree-$k$ node. If we decompose the effective degree
(c.f.\ Sect. \ref{ssec:ED}) of a node, $k$, into its susceptible and infected
neighbours such that $s+i=k$ then Table \ref{tab:SItris} enumerates the
expected number of triangles involving different states, given no correlations
between states.

$[ISS_{s,i}]_{\triangle} = \phi is [S_{s,i}]$ as the clustering coefficient,
$\phi$, gives the expected ratio of edges connecting a node's neighbours
together to the maximum possible number of such edges (Table~\ref{tab:SItris}).
Correlations between the states cannot be ignored, therefore the correlation
between nodes of state A and B, $\mathcal{C}_{AB} = \frac{N}{ \langle k
\rangle} \frac{[AB]}{[A][B]}$, is introduced which is used to account for how
many $[AB]$ pairs there are compared to how many one would expect from random
mixing, $\langle k \rangle [A]\frac{[B]}{N}$. This yields the closure equation
(\ref{eqn:inS}), and {mutatis mutandis} (\ref{eqn:inI}) is
obtained.

The original ED closure for the $\wedge$ terms must now be
modified to account for the infection events that happen within the $\triangle$
closure. The new closure is achieved by taking the original expressions and,
for $A \in \{S,I\}$, using the identity $[ISA]_{\triangle} + [ISA]_{\wedge} =
[ISA]$ along with the original and the $\triangle$ closure. For example, 
\begin{align}
  [ISS]\; =&  \sum_{j,l} [IS_{j,l}S] =\sum_{j,l} jl [S_{j,l}]\; , \\
  [ISS]_{\triangle} =& \sum_{j,l} [IS_{j,l}S]_{\triangle} = \sum_{j,l} \phi \; \mathcal{C}_{SI} jl [S_{j,l}]
\; ,\\
  \implies [ISS]_{\wedge} =&  \sum_{j,l} [IS_{j,l}S]_{\wedge} = \sum_{j,l} jl (1- \phi \; \mathcal{C}_{SI} ) [S_{j,l}] \; .
\end{align} 

Using identity (\ref{eqn:end_ident}), one gains the final clustered effective
degree closures (\ref{eqn:inS}-\ref{eqn:outI}) which when substituted into
(\ref{eqn:exactED_S}-\ref{eqn:exactED_I}) yield the clustered ED
model.
\begin{align}
  \label{eqn:inS}
  [ISS_{s,i}]_{\triangle} =&\; \phi \, \mathcal{C_{SI}} \, s i [S_{s,i}]\; , \\
  \label{eqn:outS}
  [ISS_{s,i}]_{\wedge} =&\; \frac{ \sum_{j,l}  j l (1- \phi \, \mathcal{C_{SI}} ) [S_{j,l}] }{ \sum_{j,l} j [S_{j,l}]} s [S_{s,i}]\; ,\\
  \label{eqn:inI}
  [ISI_{s,i}]_{\triangle} =&\; \phi \, \mathcal{C_{II}} \, s i [I_{s,i}]\; , \\
  \label{eqn:outI}
  [ISI_{s,i}]_{\wedge} =&\; \frac{ \sum_{j,l}  l(l-1) \big(1 - \phi \, \mathcal{C_{II}} \big) [S_{j,l}] }{ \sum_{j,l} j [I_{j,l}]} s [I_{s,i}] \; . 
\end{align}

\clearpage
\section{Simulation Study\label{sec:sim_stud}}

To test the accuracy of the clustered ED model we perform a
simulation study.

\subsection{Methodology\label{ssec:methodology}}

First, unclustered networks are generated for three different degree
distributions according to the methods presented in Sect. \ref{sec:net_gen}.
The big-V algorithm (Fig. \ref{fig:bigV} \ding{178}) was then applied to each
unclustered network to generate a series of networks with approximate
clustering coefficients $\phi \in \{0.05, 0.10, 0.15, 0.20, 0.25, 0.30\}$.

Numerical simulation was performed using an individual-based analogue of
Gillespie's algorithm \cite{gillespie1976general}, which generates a
statistically correct trajectory with regard to the master equation of the
underlying stochastic process.  Simulations were performed for 20 uniquely
generated networks of $10^5$ nodes with the dynamics being simulated on each
network 5 times, for a total of 100 epidemics per network type. To account for
the large stochastic variability at the beginning of an epidemic, we shifted
the time-origins of each of the 100 epidemics to coincide at the point where
500 individuals are infected before averaging the dynamics, which we then
compare to each differential equation model.

We choose to generate and simulate networks of mean degree four as a rigorous
model comparator---closures perform better as $\langle k \rangle$ increases as
the behaviour tends towards the simple SIR model since these networks are
closer to complete graphs. 

\subsection{Results}

\begin{figure}[H]
\centering
\includegraphics[width=\textwidth]{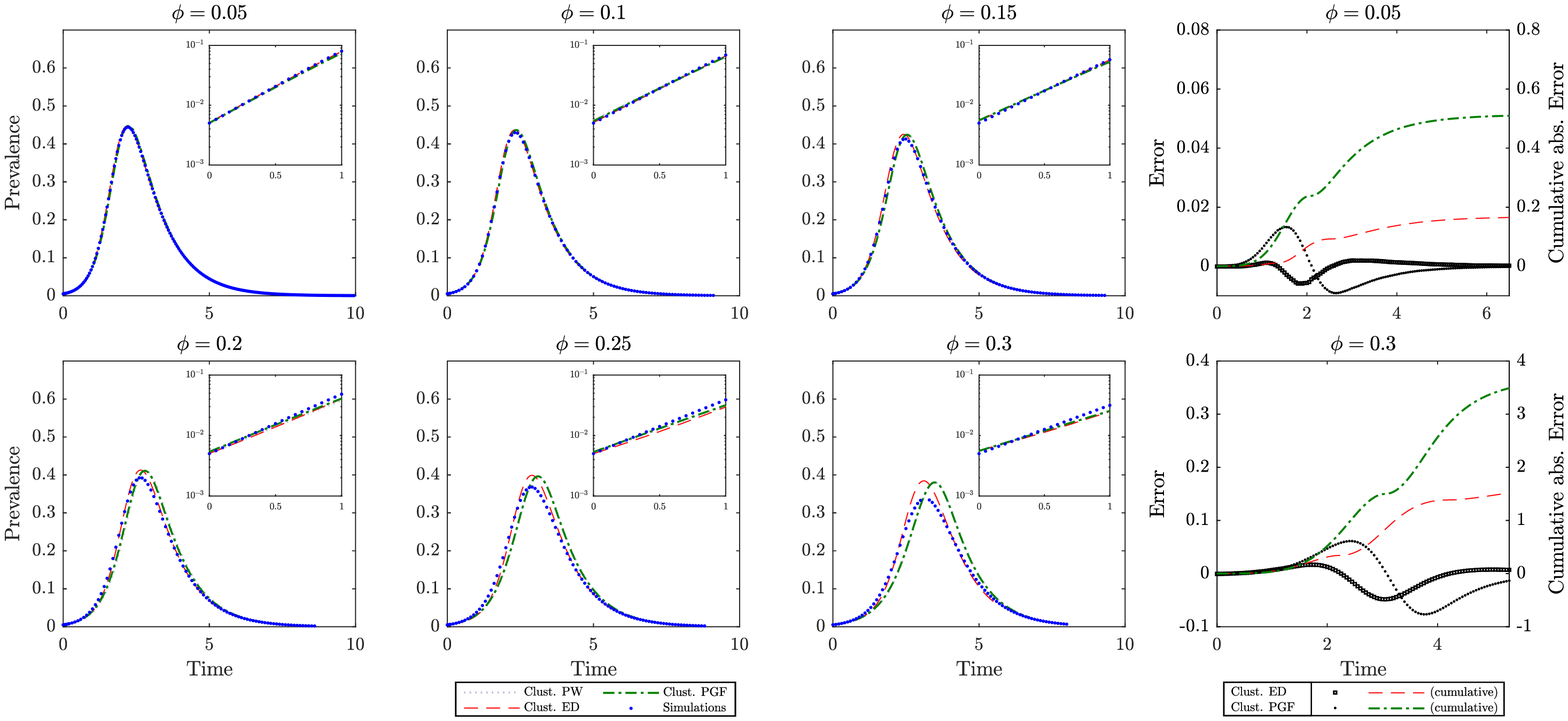}
\caption{Simulated epidemic dynamics for 4-regular networks with $\gamma=1, \tau=2, N=10^5$.}
\label{fig:reg}
\end{figure}

\begin{figure}[H]
\centering
\includegraphics[width=\textwidth]{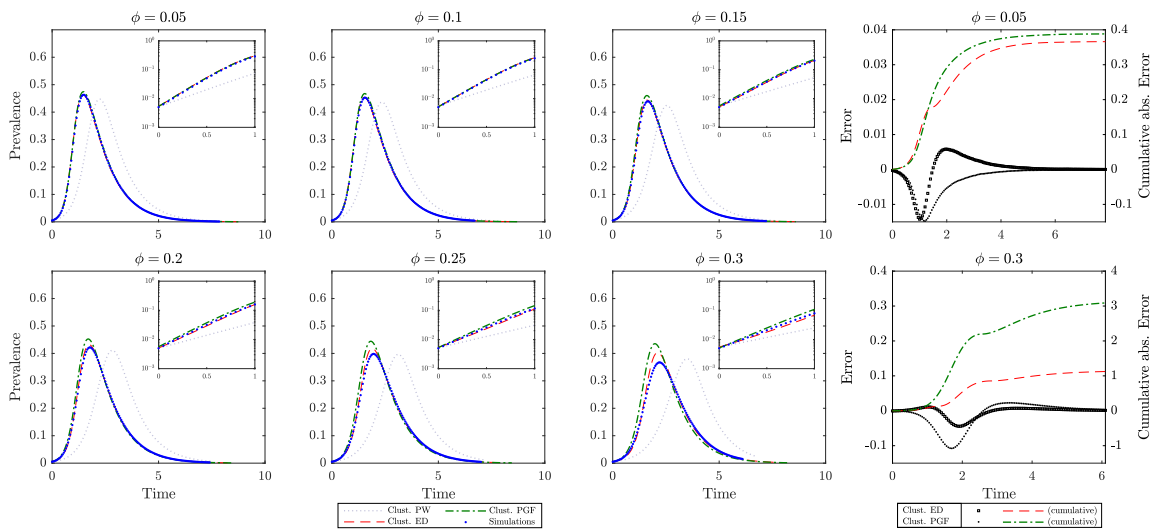}
\caption{Simulated epidemic dynamics for Erd\H{o}s-R\'enyi networks with
$\langle k \rangle = 4, \gamma=1, \tau=2, N=10^5$.}
\label{fig:pois}
\end{figure}

\begin{figure}[H]
\centering
\includegraphics[width=\textwidth]{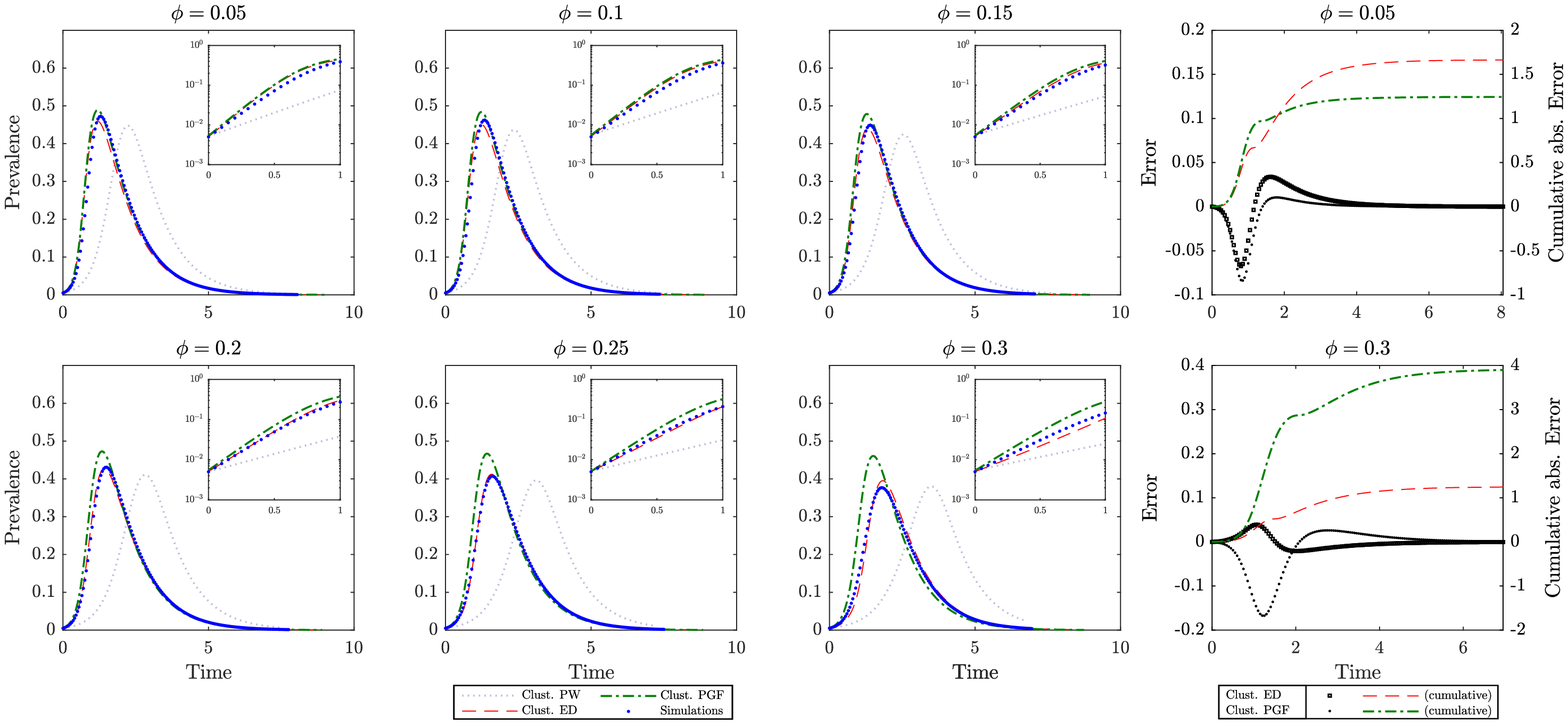}
\caption{Simulated epidemic dynamics for Negative binomial networks with
$\langle k \rangle = 4, r=5, \gamma=1, \tau=2, N=10^5$. We use the formulation
  where we are counting $k$ successes given $r$ failures.}
\label{fig:nb}
\end{figure}

The results in Figs.~\ref{fig:reg}--\ref{fig:nb} show that
our novel clustered ED model generally outperforms the clustered
PGF approach in terms of errors in expected prevalence, with exceptions to this
only occuring after the epidemic peak.

\begin{figure}[H]
\begin{center}
\includegraphics[width=0.99\textwidth]{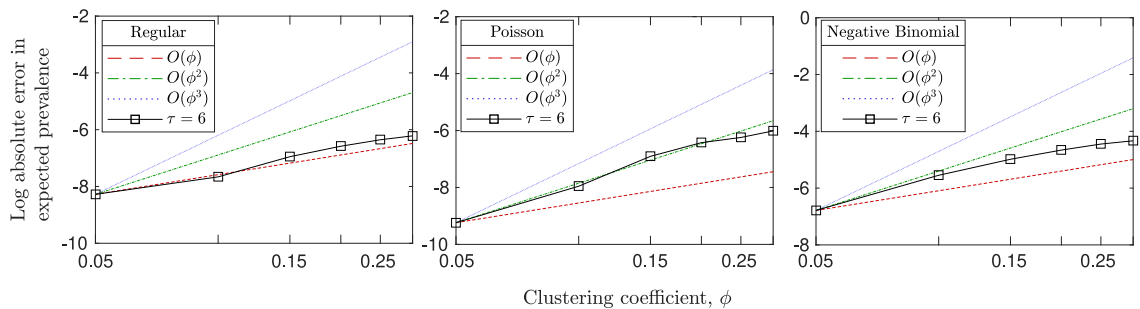}
  \caption{Error between the clustered ED model and expected prevalence of the stochastic simulations detailed in Sec. \ref{ssec:methodology} is plotted against errors of order $\phi$, $\phi^2$, and $\phi^3$ plotted in red, green, and blue respectively.}
\label{fig:err}
\end{center}
\end{figure}

Work by Pellis et al. \cite{Pellis:2015} argued (based on a rigorous analysis of finite-size
networks) that we should expect moment closure to be exact when (i) triangles
are not overlapping and (ii) recovery happens after a constant time (or not at
all as when $\gamma=0$). When the clustering coefficient is small, the
proportion of overlapping triangles will be $O(\phi^2)$, and so for small
values of $\gamma$ we expect errors in the prediction of prevalence of
infection to be $O(\phi^2)$, while for larger values we expect them to be
$O(\phi)$.  We found that in general, obtaining accurate assessments of
absolute error in predictions of prevalence was numerically challenging, and it
is likely that deeper theoretical understanding of the source of errors would
be required to perform a definitive computational analysis of this question.
Nevertheless, we were able to obtain the results shown in Fig.~\ref{fig:err},
which demonstrate that as expected errors lie between the $O(\phi)$ and
$O(\phi^2)$ lines.

\section{Discussion}

In this paper, we have shown how to combine effective degree approaches to
epidemics on heterogeneous networks with moment closure approaches to
clustering. Numerical results suggest that the errors introduced in so doing
may be better than $O(\phi)$ but worse than $O(\phi^2)$, meaning that the
potential for improvements remains. In particular, the closure due to
Kirkwood \cite{kirkwood1942} could be replaced by the improved closure in
\citep{housetracing} or the maximum entropy method \cite{Rogers:2011}.  In
particular, in the large $\phi$ limit we expect graphs to be dominated by
cliques \citep{DelGenio:2013}, whose differential equation limit is described
by equations such as those written in \cite{Ball:1999}.

\end{document}